\documentclass[aps]{revtex4}
\usepackage{graphicx,psfrag}
\def\no{\noindent}
\def\bc{\begin{center}}
\def\ec{\end{center}}

\def\beq{\begin{equation}}
\def\eeq{\end{equation}}

\def\bj{{\bf j}}

\def\br{{\bf r}}
\def\bq{{\bf q}}
\def\bk{{\bf k}}

\def\bE{{\bf E}}

\begin{document}

\title{
Random edge states on a finite lattice 
}

\author{K. Ziegler}
\affiliation{Institut f\"ur Physik, Universit\"at Augsburg\\
D-86135 Augsburg, Germany\\
}
\date{\today}

\begin{abstract}
A finite photonic lattice with two bands and a random gap is considered. Using
a two-dimensional Dirac equation, the effect of a random sign of the Dirac mass
is studied numerically. The edge state at the sample
boundary has a strong influence on the electromagnetic field and its polarization
inside the sample. The creation of edge states through a randomly fluctuating 
sign of the Dirac mass defeats Anderson localization and allows the 
electromagnetic field to distribute over the entire sample. The width of the
distribution increases with an increasing gap due to increasing sharpness of
the edge states. These results are compared with those of a random 
one-band Helmholtz equation. In contrast to the Dirac model, the one-band
model displays a clear signature of Anderson localization. 
\end{abstract}

\maketitle

\section{Introduction}

In finite gapped systems the boundary plays a crucial non-trivial role due to the 
formation of edge states. This implies that a simple (e.g., plane-wave) ansatz for its
solution is insufficient, since the edge states decay exponentially
on a scale which is inversely proportional to the gap.

For the subsequent study it is essential that the field equation depends on a real gap
parameter, whose sign can be changed. This requires a system consisting of two bands with
degeneracy points (spectral nodes). A typical example is a photonic crystal with a bipartite lattice
structure such as a honeycomb lattice. Many other examples have been discussed in the 
literature~[\onlinecite{maradudin91,peleg07,haldane08,raghu08,zhang08,wang09,ochiai09,ablowitz09,
zandbergen10,huang11,bravo12,fefferman12,rechtsman13,keil13,ma15,cheng16}].
Without considering lattice details we focus on the
vicinity of a generic Dirac node, where the spectrum is linear. 

Edge states in a gapped two-dimensional electron gas cause the Quantum-Hall Effect, which is
experimentally observed as a quantized Hall conductivity~[\onlinecite{klitzing80}]. 
The edge states themselves are not
directly accessible, though. The situation is quite different in photonic systems, where the
electromagnetic field, which is the formal analog of a electronic single-particle quantum
state, is directly observable. The tremendous progress in the design of
photonic metamaterials during recent years made it possible to create and observe edge 
states experimentally and to measure their topological properties, such as their Berry
curvature.
Another advantage of photonics in comparison to electronics is that photons are not charged
and that they do not interact directly with each other. Moreover, photons cover a wide range
of length scales which are described by the same Maxwell theory. This provides the 
opportunity to design metamaterials from nano-- to centimeter length scales with the same
type of physical properties. For example, there have been similar experiments with 
visible light 
as well as with microwaves in the frequency regime down to 10 GHz.
Thus, the characteristic lengths in the photonic experiments vary over seven orders of 
magitude from $10^{-9}$m ... $10^{-2}$m. 

Besides the edge state at the boundary of the photonic crystal, edge states can also be 
created inside a photonic crystal by an inhomogeneous gap parameter~[\onlinecite{haldane08,raghu08}].
This effect was studied recently in several experiments with dielectric as well as 
with metallic samples~[\onlinecite{keil13,cheng16}], where edge states were directly identified
between regions with different signs of the Dirac mass.

The robustness of the edge states may also lead to a robust behavior of photons even in
the presence of a random distribution of edges. This effect was analyzed in an infinite
system with the result that photons propagate in this case in the form of ray modes and 
avoid Anderson localization~[\onlinecite{1404,ziegler17a,ziegler17b}]. One implication
is that photons from a central source inside a finite crystal can propagate to the 
boundary and create an edge state there.

In the following we analyze edge states in photonic crystals with two bands and nodal 
spectral degeneracies. The two-component electromagnetic field in the $x$--$y$ plane 
is a solution of the two-dimensional (2D) Dirac equation and represents formally a Dirac
spinor. 
This work was inspired by a series of recent experiments with edge states in photonic 
crystals near a slightly gapped Dirac node~[\onlinecite{huang11,rechtsman13,ma15,cheng16}].
Here the aim is to study properties of the edge states caused by a random Dirac mass
and to propose its characterization through a spatial distribution of the intensity.
In particular, the transfer of field intensity from a source at the center of the sample to
its boundary will be studied.

\begin{figure*}[t]
\includegraphics[width=7.5cm,height=6.5cm]{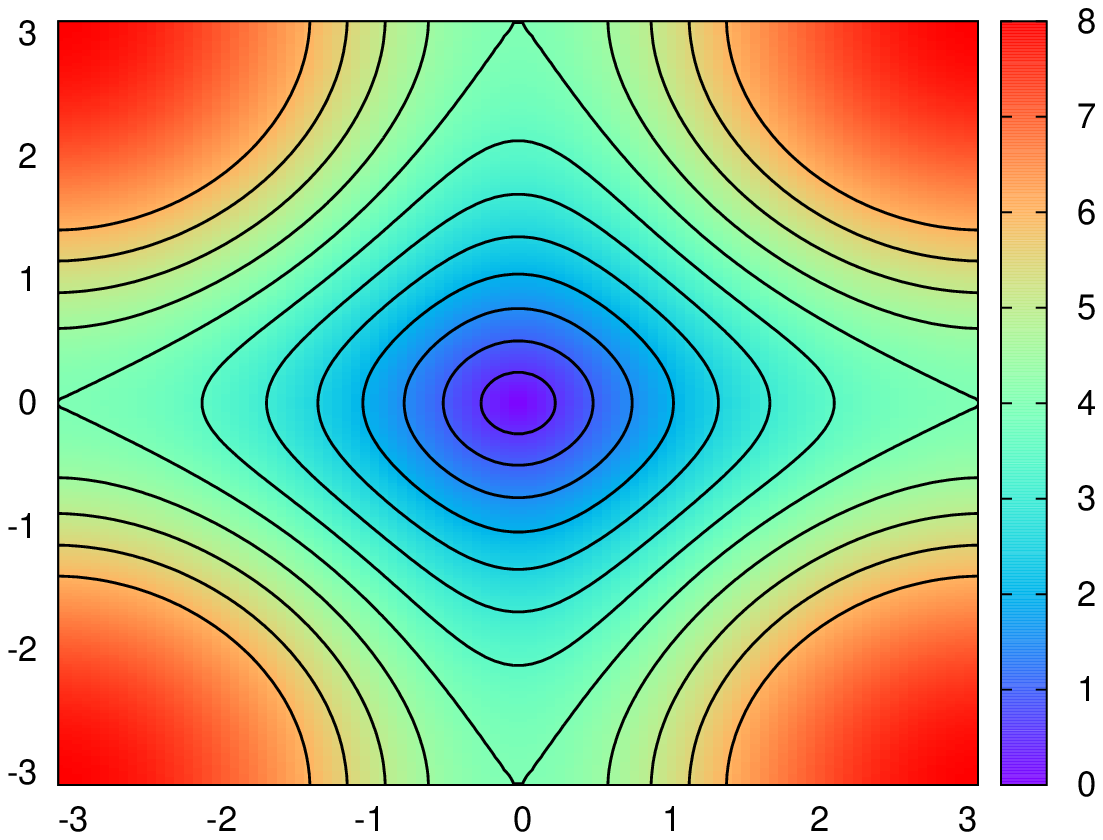}
\includegraphics[width=7.5cm,height=6.5cm]{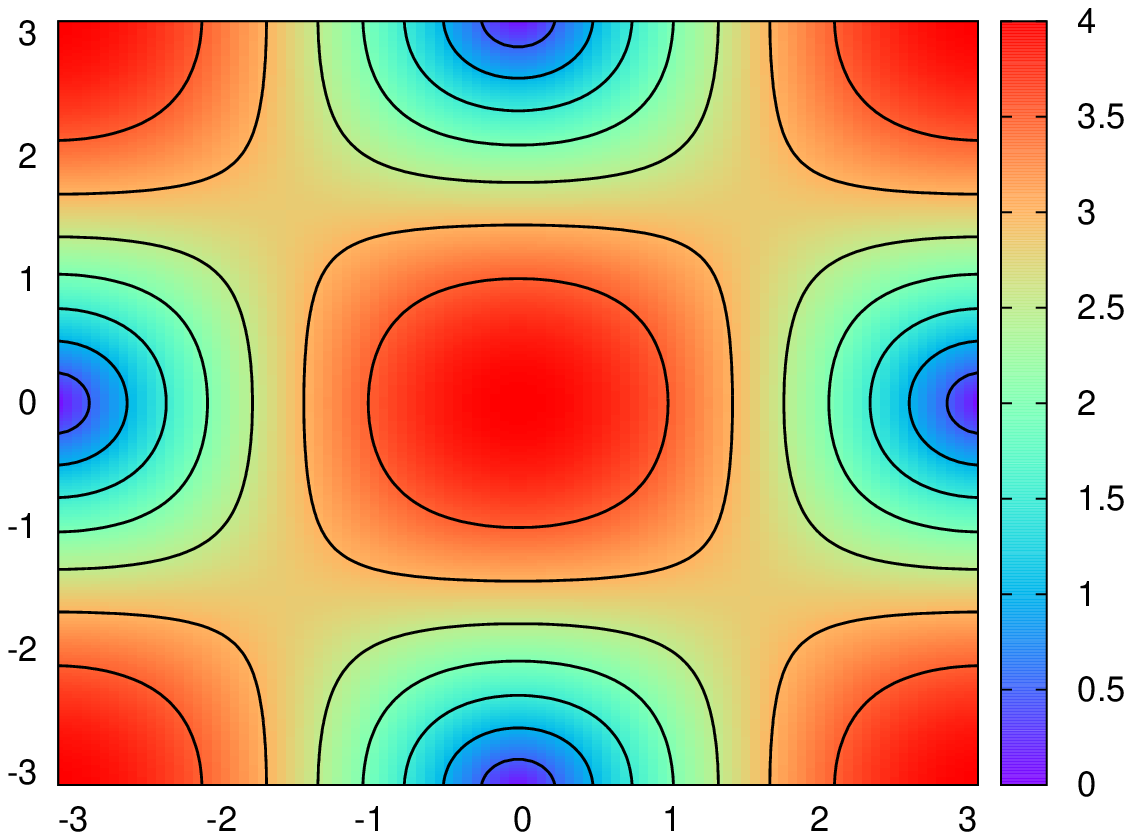}
\caption{
The $k_x$--$k_y$ dispersion of an infinite system with periodic boundaries from Eq. (\ref{hill00}) 
for $\alpha=1$ and with a uniform Dirac mass $m=0$ (left) and $m=4$ 
(right). $m=8$ is another case of a vanishing gap at the corners $k_x=\pm\pi$, $k_y=\pm\pi$.
For all other values of $m$ the spectrum has a gap.
}
\label{fig:disperse1}
\end{figure*}

\section{The electromagnetic field in a photonic crystal}

The fundamental Maxwell equations of the three-component electric (${\bf E}$) and the 
three-component magnetic (${\bf H}$) field
\beq
{\bf \nabla}\times {\bf H}
=\frac{\epsilon}{c}\frac{\partial{\bf E}}{\partial t}+\frac{4\pi}{c}{\bf j}
\ , \ \ \ 
{\bf \nabla}\times {\bf E}
=-\frac{1}{c}\frac{\partial{\bf H}}{\partial t}
\label{maxwell00}
\eeq
with the dielectric coefficient $\epsilon$ and speed of light $c$ imply the equation for 
${\bf E}$
\beq
{\bf \nabla}\times{\bf \nabla}\times {\bf E}
=-\frac{1}{c}\frac{\partial}{\partial t}\left[\frac{\epsilon}{c}
\frac{\partial{\bf E}}{\partial t}+\frac{4\pi}{c}{\bf j}\right]
\ .
\label{maxwell1}
\eeq
After a Fourier transformation $t\to\omega$ we obtain for the frequency-dependent Fourier components 
${\bf E}(\br,\omega)$ the inhomogeneous differential equation 
\beq
-{\bf \nabla}\times{\bf \nabla}\times {\bf E}+\frac{\epsilon \omega^2}{c^2}{\bf E}
=\frac{4i\omega\pi}{c}{\bf j}(\omega)
\ ,
\label{maxwell2}
\eeq
where ${\bf j}(\omega)$ is the Fourier transform of the current density.
The factor $\epsilon \omega^2/c^2$ describes the effect of a dielectric medium. 

Next we consider a photonic crystal which is characterized by a periodic space-dependent dielectric tensor 
$\epsilon$ in the $x$--$y$ plane.
This implies that the spectrum of the operator acting on $\bE$ 
has a band structure~[\onlinecite{maradudin91,john87,bykov72,ohtaka87,yablonovitch87}]. 
It was argued that for a triangular or hexagonal arrangement of dielectric rods or cylinders 
there are two spectral nodes in the band structure \cite{maradudin91,haldane08}. 
Near these spectral nodes Eq. (\ref{maxwell2})
can be reduced for the transverse electric (TE) mode ${\bf E}=(E_1,E_2,0)$, ${\bf H}=(0,0,H_3)$
to two inhomogeneous two-dimensional Dirac equations for the two nodes
\beq
H_{D;\pm}{\bf E}_\pm=4ic\pi{\bf j}(\omega)
\label{dirac00}
\eeq
with
\beq
H_{D;\pm}=v_D\pmatrix{
M & i\partial_x\pm\partial_y \cr
i\partial_x\mp\partial_y & -M \cr
}
\eeq
for the two-component field ${\bf E}_\pm=(E_{\pm,1},E_{\pm,2})$ with the Dirac 
mass $M$ and the Dirac velocity $v_D$.
Furthermore, the degeneracy of the two Dirac equations can be
lifted by an asymmetric gap opening when, for instance, the
Faraday effect is included for the medium~[\onlinecite{haldane08}] or by an
asymmetric design of microwave metamaterials~[\onlinecite{cheng16}] .
For the electric component of the TE mode in the $x$-$y$ plane
at the proper frequency $\omega_D$ there is only one Dirac photon left \cite{haldane08}:
\beq
H_D{\bf E}
=4ic\pi{\bf j}(\omega)
\label{dirac01}
\eeq
with $H_D\equiv H_{D;+}$.
In order to create finite bands, as observed in photonic metamaterials, 
the Dirac equation must be discretized in space, reflecting the lattice structure
of the photonic crystal.
However, the naive discretization by replacing the differential operators $\partial_{x,y}$ 
with difference operators leads to additional Dirac nodes, an effect which is called fermion
doubling or fermion multiplication in lattice gauge theory~\cite{Susskind1977}. There
are modifications of the Dirac operator, though, to circumvent this 
problem~\cite{Stacey1982,ziegler96,beenakker08,beenakker10}.
Here we adopt an idea of Susskind and replace $H_D$ by the lattice Dirac operator~[\onlinecite{hill14}]
\beq
H_{DD}=\frac{v_D}{a}\pmatrix{
m +\alpha(\Delta_1+\Delta_2-4) & id_1+d_2 \cr
id_1-d_2 & -m-\alpha(\Delta_1+\Delta_2-4) \cr
}
\label{lattice_dirac}
\eeq
with the lattice difference operators 
\beq
\Delta_{j}(\br\br')=\cases{
1 & for $|x_j-x_j'|=a$ \cr
0 & otherwise \cr
} \ , \ \
d_{j}(\br\br')=\cases{
1 & for $x_j-x_j'=-a$ \cr
-1 & for $x_j-x_j'=a$ \cr
0 & otherwise \cr
}
\eeq
for 
the lattice constant $a$ and with the Dirac mass $m=aM$. $m$ is now a dimensionless 
parameter which measures an inverse length in units of $1/a$.
This lattice operator is generic for a larger class of discrete photonic metamaterials
which possess Dirac nodes.
For the special case of an infinite lattice and a uniform Dirac mass the Fourier representation 
of $H_{DD}$
\begin{equation}
{\tilde H}_{DD}= \frac{v_D}{a}\pmatrix{
m+2\alpha(\cos k_1+\cos k_2-2)  & 2\sin k_1 - i2\sin k_2 \cr
2\sin k_1 + i2\sin k_2 & -m- 2\alpha(\cos k_1+\cos k_2-2) \cr
}
\ \ \
(-\pi\le k_j <\pi)
\label{main_ham}
\end{equation}
provides the two-band dispersion
\beq
E_\bk=\pm\frac{v_D}{a}\sqrt{[m-4\alpha+2\alpha(\cos k_1+\cos k_2)]^2+4\sin^2k_1+4\sin^2k_2}
\ .
\label{hill00}
\eeq
There are two interesting values for the parameter $\alpha$:
for $\alpha=1$ there is only one node at $\bk=0$ if $m=0$ and another node for $m=8$ with $k_x=\pm\pi$
and  $k_y=\pm\pi$.
A gap closing with two nodes exists if $m=4$ (cf. Fig. \ref{fig:disperse1}).
A simpler case is $\alpha=0$, where we have four nodes if $m=0$ and always a gap if $m\ne 0$.

In the following we consider the Dirac Eq. (\ref{dirac01}) with the lattice Dirac operator $H_{DD}$ 
for $\alpha=1$ from a local source at $\br_0$. For simplicity, the special choice $\bj=\bj_0\delta_{\br,\br_0}$ 
with $\bj_0=(j,0)$ is used subsequently.

\subsection{Edge states}

\begin{figure*}[t]
\includegraphics[width=7cm,height=7cm]{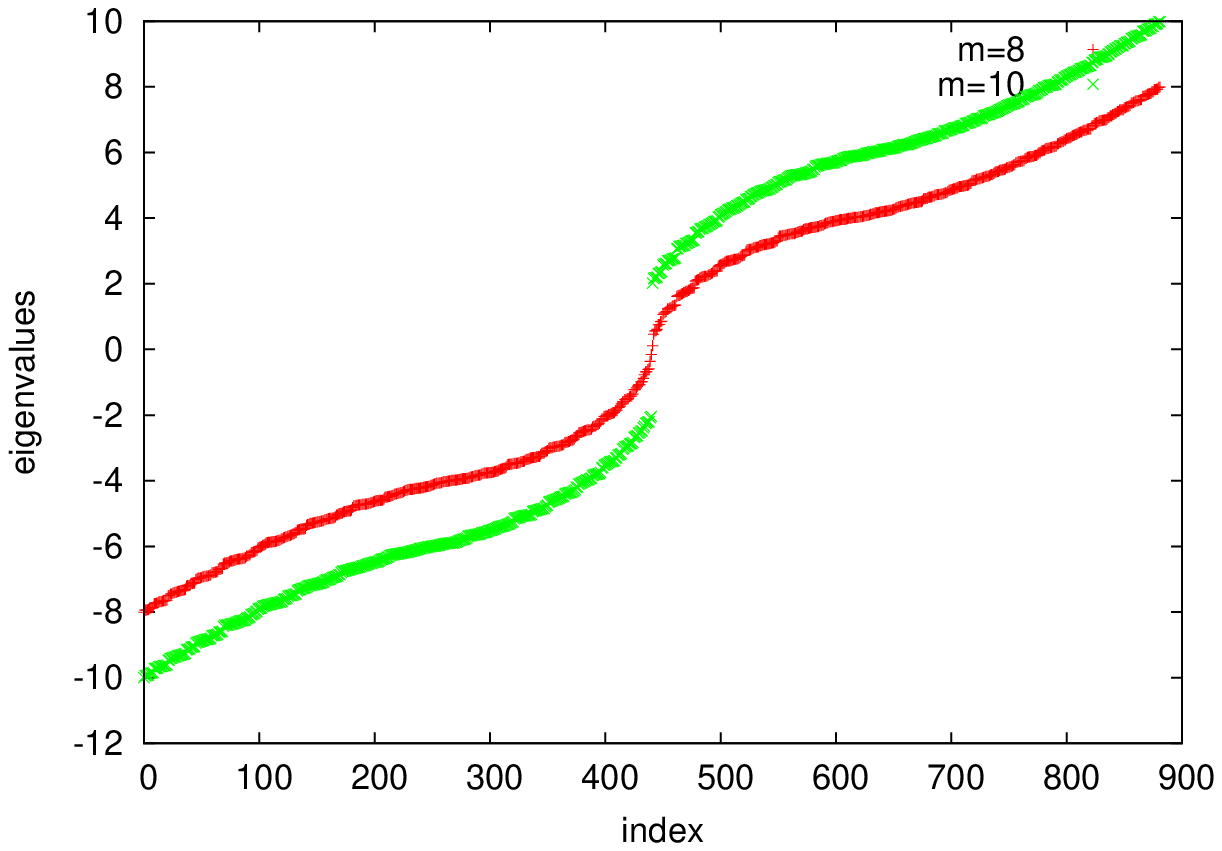}
\includegraphics[width=7cm,height=7cm]{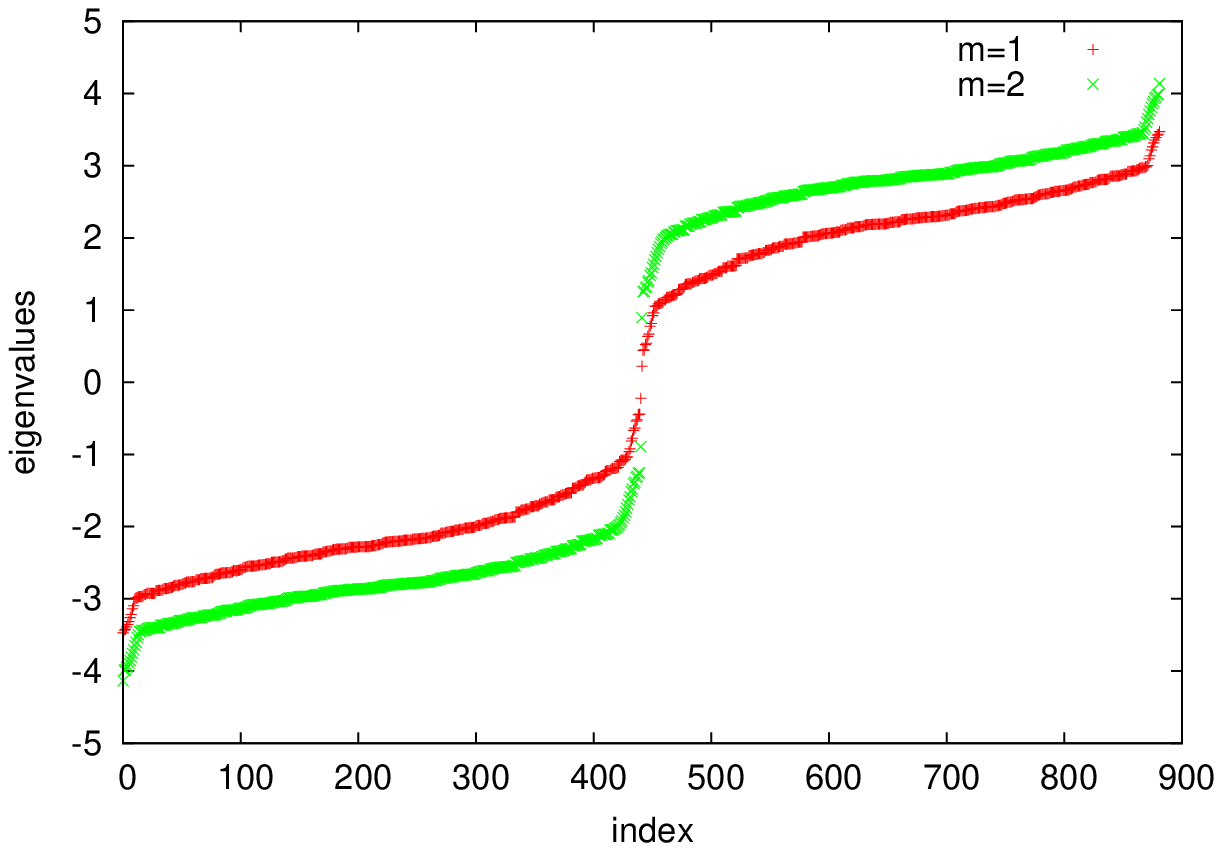}
\caption{
The eigenvalues for different uniform gaps in the case of one (left) and four (right) nodes
for a lattice of size $21\times 21$.
The gaps are filled with edge states up to a critical gap value. These critical values
are $m_c\approx 9$ for one node (i.e., $\alpha=1$) and $m_c\approx 1$ for four nodes (i.e., $\alpha=0$). 
}
\label{fig:eigenvalues}
\end{figure*}

Edge states appear when the gap is not larger than the width of the bands. This can be
easily seen in the simple example of a spinor plane wave with wavevector $\bk$ and the 
dispersion $E_\bk=\pm\sqrt{m^2+k^2}$. Then an edge state appears for an imaginary 
wavevector $\bk=i\bq$ ($\bq$ real), which has the dispersion
\[
E_\bk=\pm\sqrt{m^2-q^2}
\ .
\]
If the band restricts the wavevectors to $k\le \lambda$, there are no edge states at zero energy
if $m^2>\lambda^2$. Thus, for the existence of an edge state we need a Dirac
mass smaller than the width of a single band.
In the case of the lattice Dirac operator (\ref{lattice_dirac}) with uniform Dirac mass
on a finite lattice we get a critical Dirac mass, as depicted in Fig. \ref{fig:eigenvalues}, 
depending on $\alpha=0,1$: For $\alpha=0$ a gap opens at zero energy in the spectrum when 
$m$ is larger than $m=1$, and for $\alpha=1$ this happens when $m$ is larger than $m=8$.

\begin{figure*}[t]
\includegraphics[width=7cm,height=7cm]{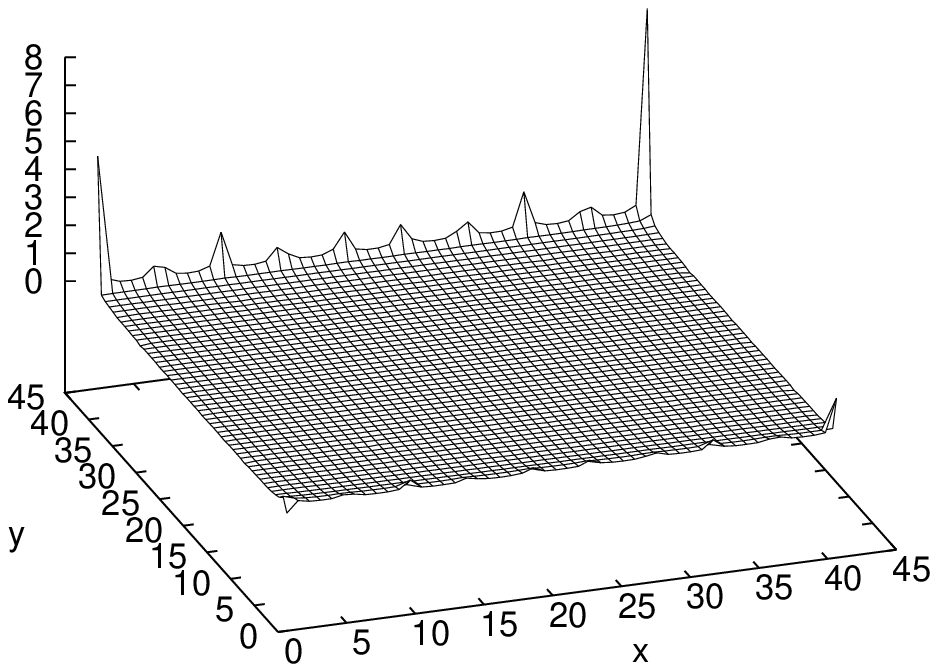}
\includegraphics[width=7cm,height=7cm]{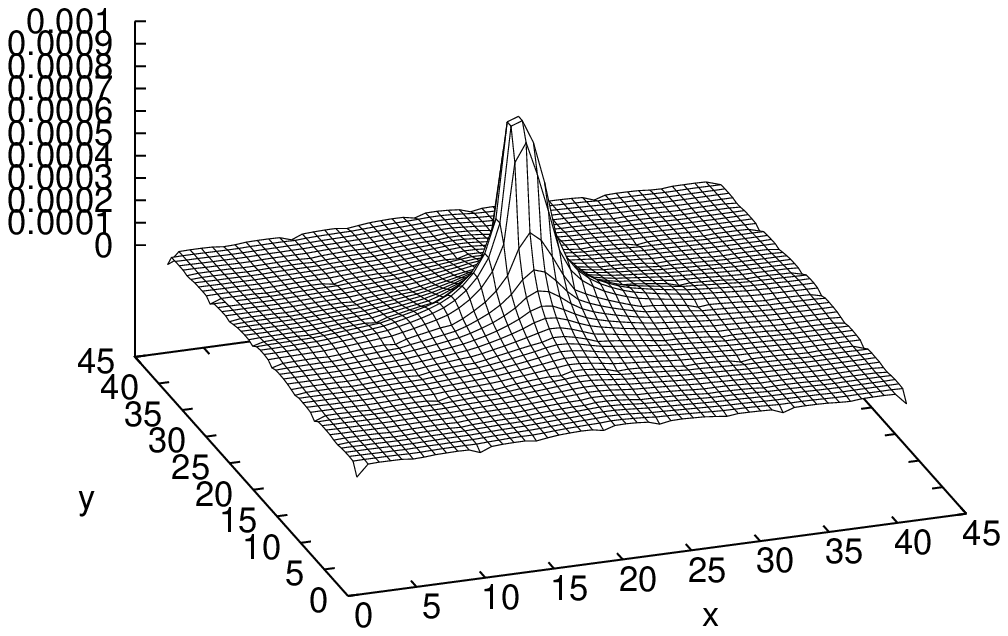}
\caption{
Intensity transfer from a source at the corner $x=1$, $y=0$ along the boundary (left) 
and the intensity localization near a central source (right) for a uniform Dirac mass $m=2$. 
}
\label{fig:uniform1}
\end{figure*}

\begin{figure*}[t]
\includegraphics[width=7cm,height=7cm]{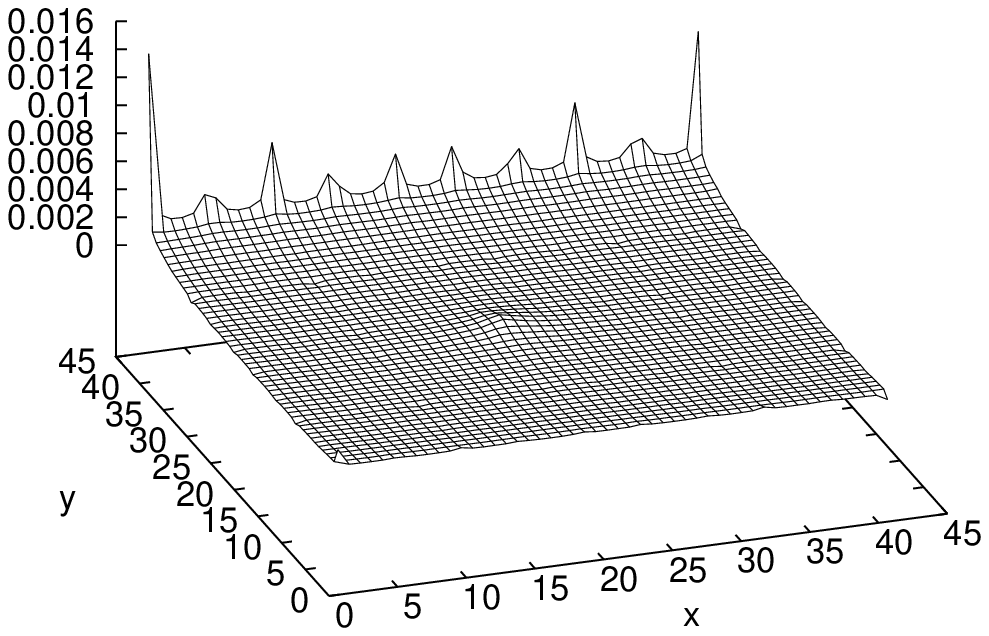}
\includegraphics[width=7cm,height=7cm]{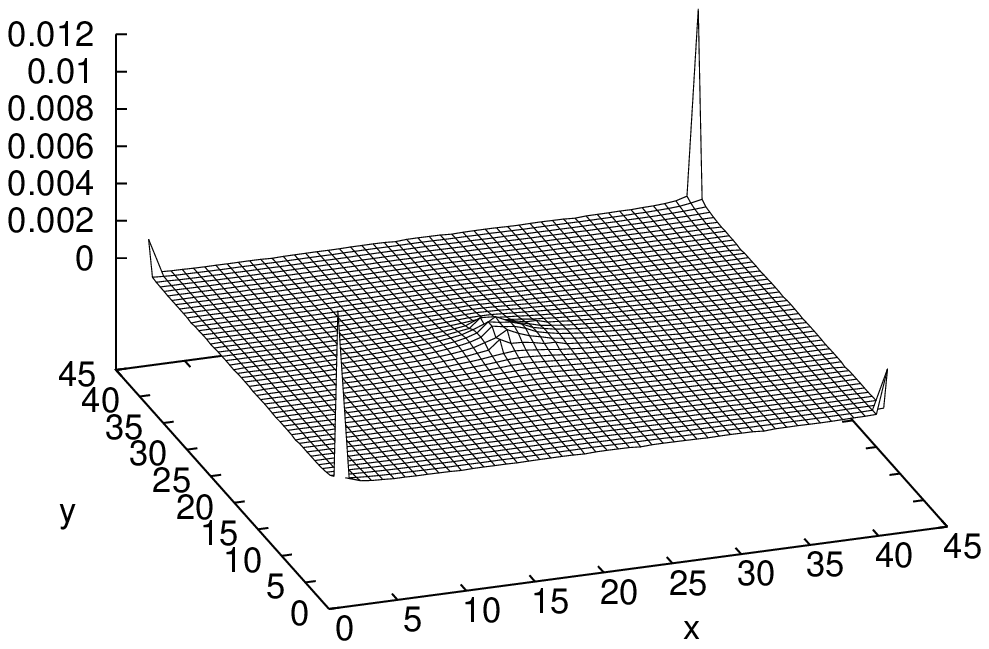}
\caption{
Intensity transfer from a central source to the boundary for a uniform Dirac mass in two 
exceptional cases with very small gap ($m\approx 0$ and $m\approx 4$). 
}
\label{fig:uniform2}
\end{figure*}

\begin{figure*}[t]
\includegraphics[width=7cm,height=7cm]{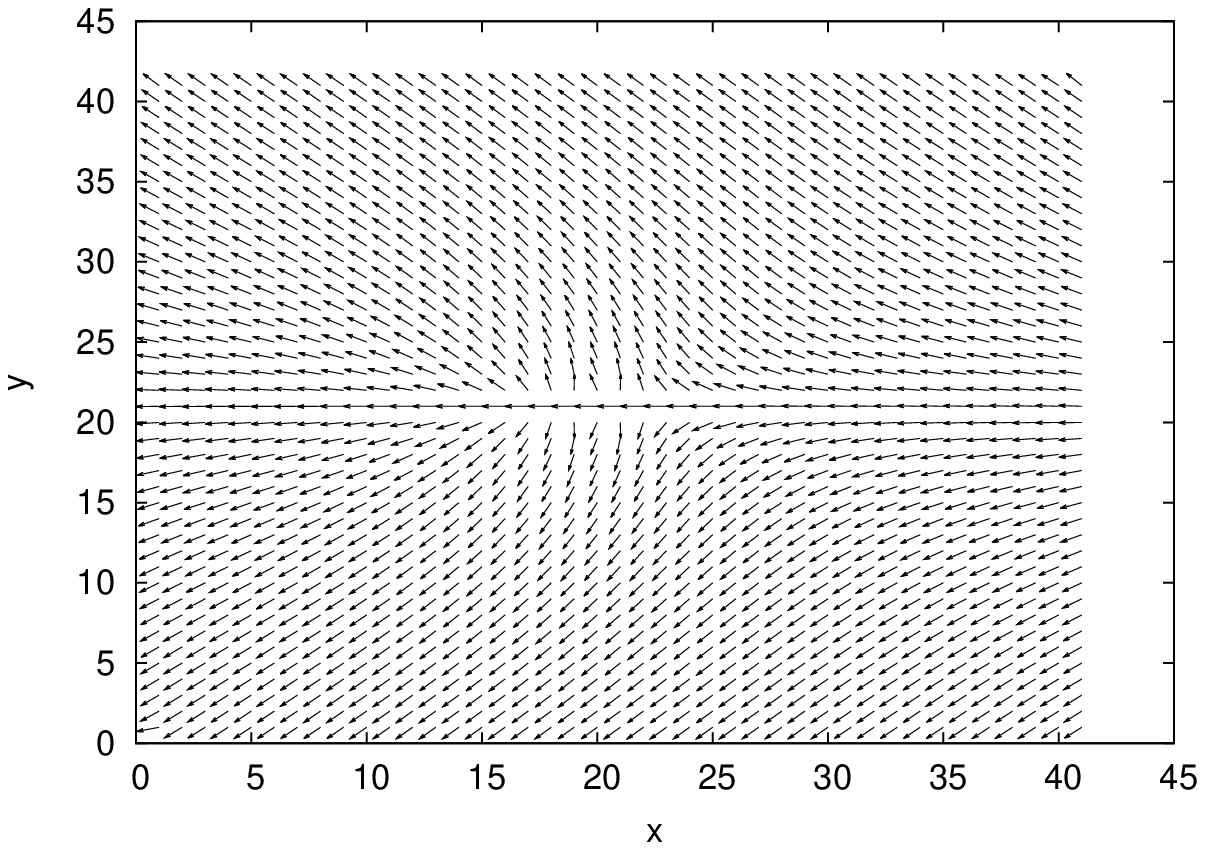}
\includegraphics[width=7cm,height=7cm]{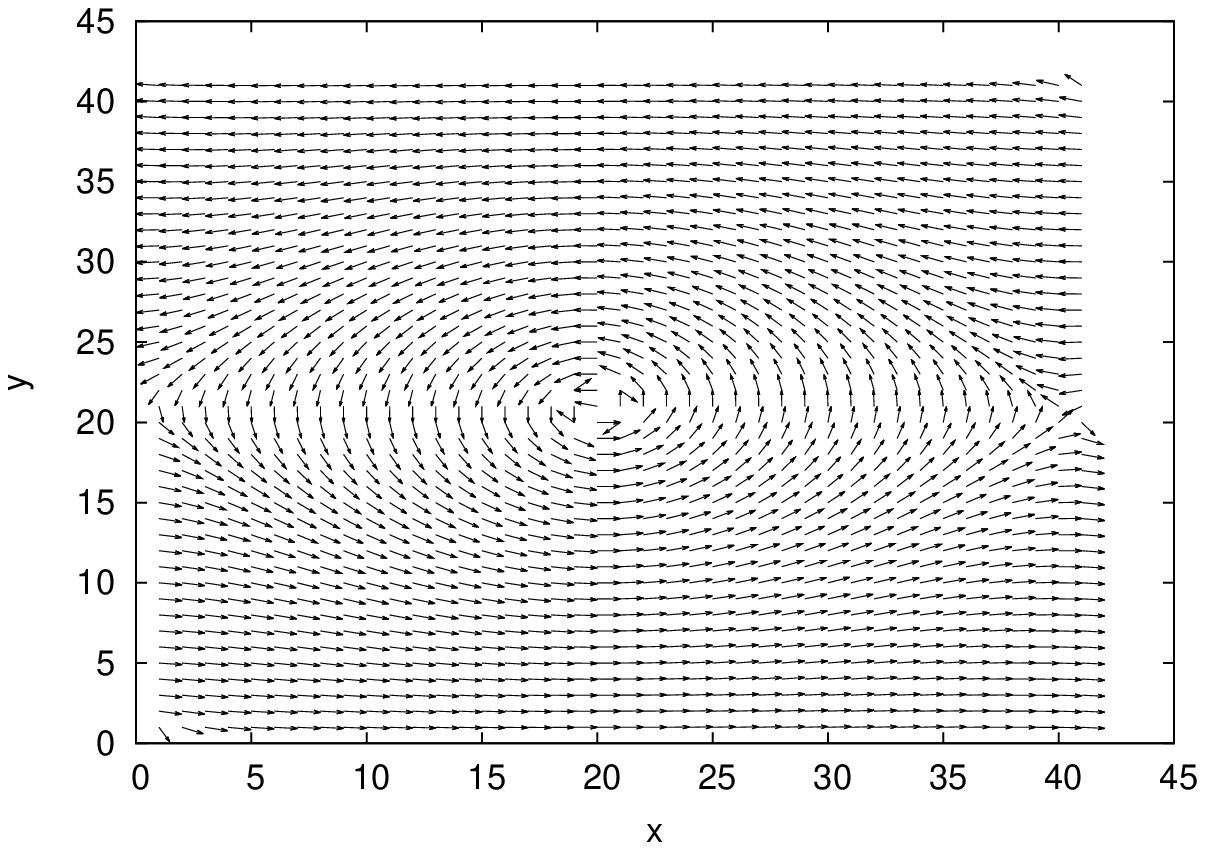}
\includegraphics[width=7cm,height=7cm]{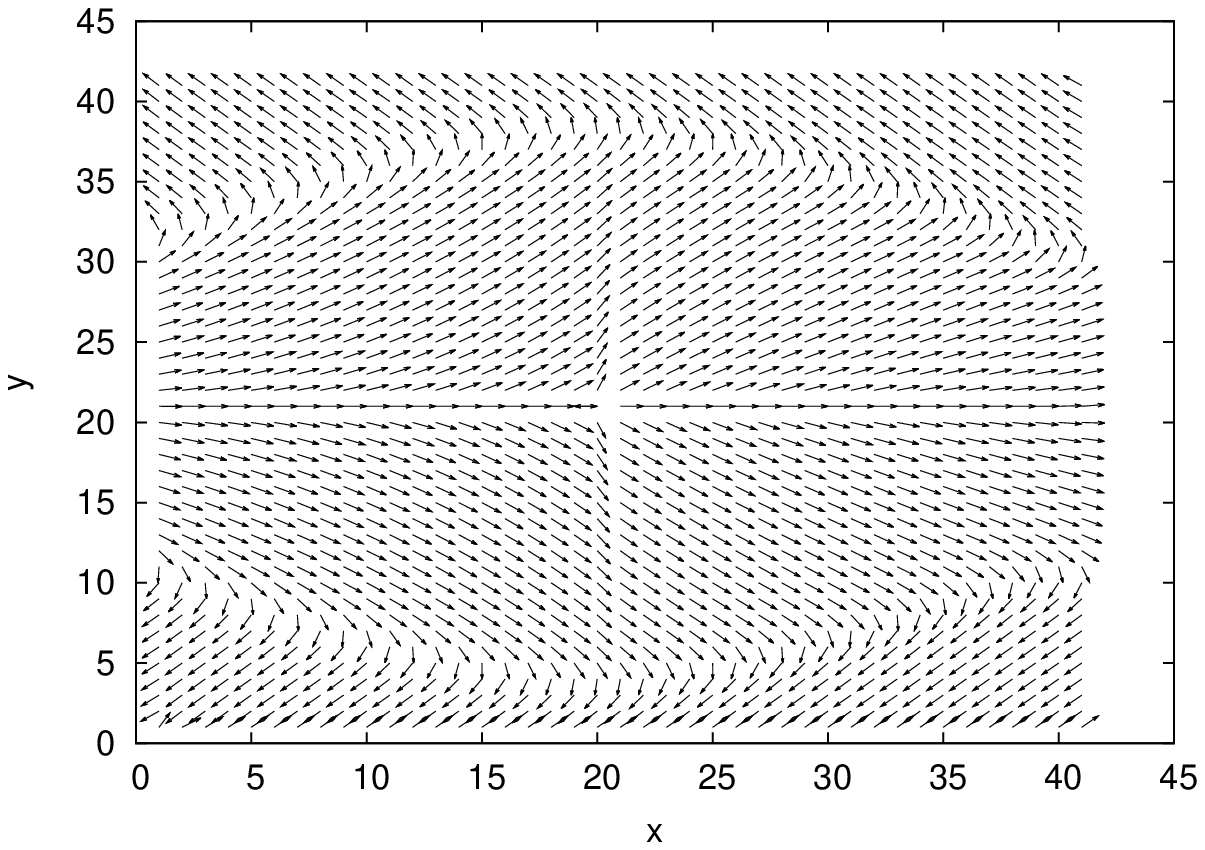}
\includegraphics[width=7cm,height=7cm]{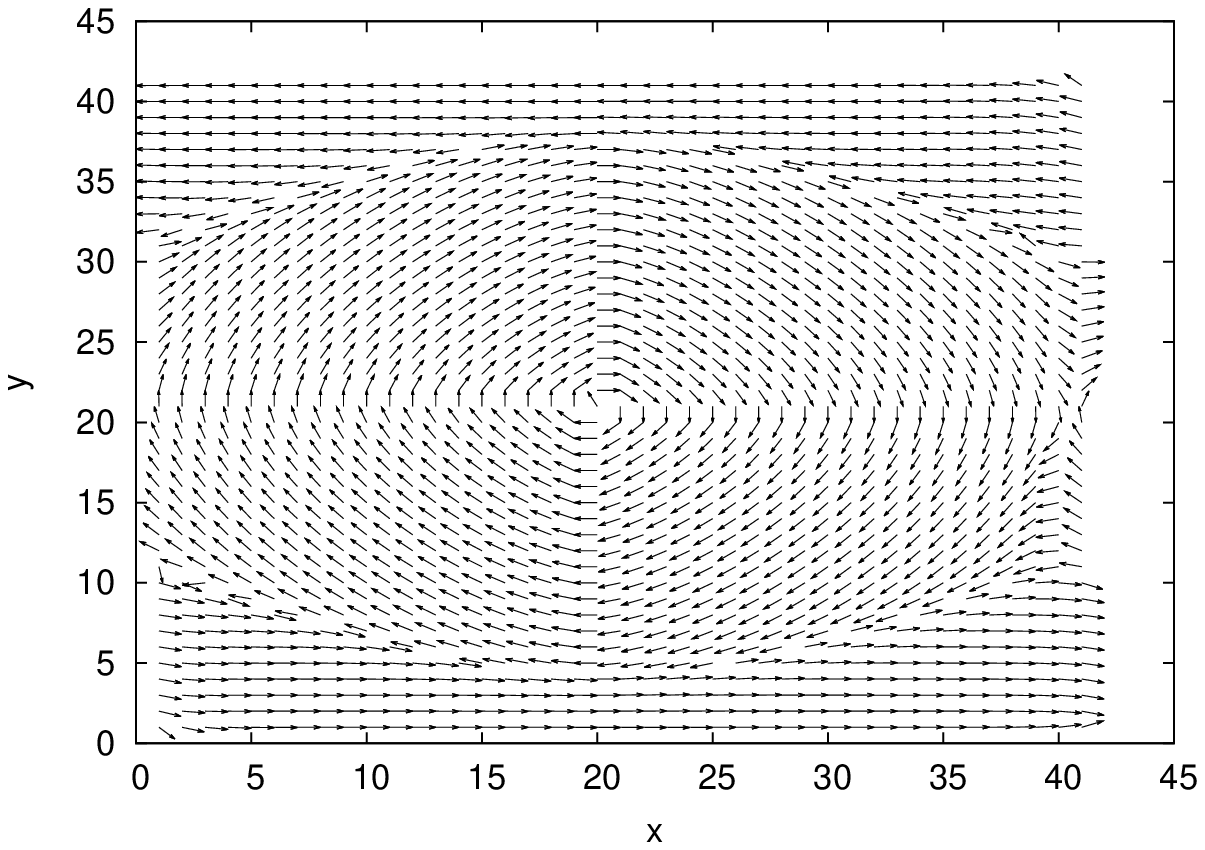}
\caption{
Effect of a sign flip of a uniform Dirac mass 
(top: $m=1$, bottom: $m=-1$) on the normalized electromagnetic field (left)
and on the polarization (right).
}
\label{fig:uniform3}
\end{figure*}
\begin{figure*}[t]
\includegraphics[width=7cm,height=7cm]{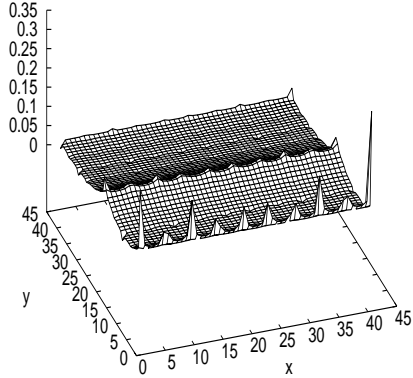}
\caption{
Intensity transfer from the central source to the boundary due to a straight edge 
along the $x$ direction in the middle of the sample with $m=0.45$ for $y<21$ 
and $m=-0.45$ for $y>21$.
}
\label{fig:edge1}
\end{figure*}

\begin{figure*}[t]
\includegraphics[width=7cm,height=7cm]{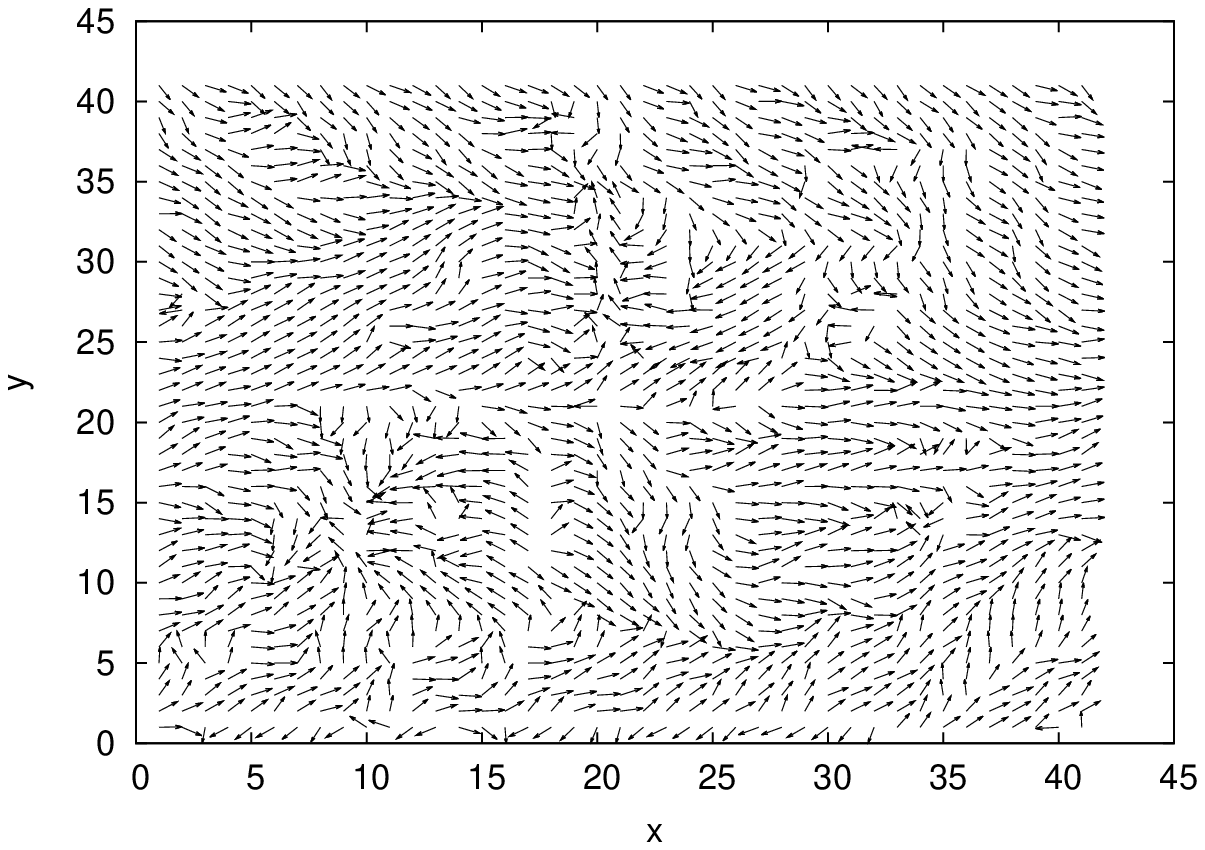}
\includegraphics[width=7cm,height=7cm]{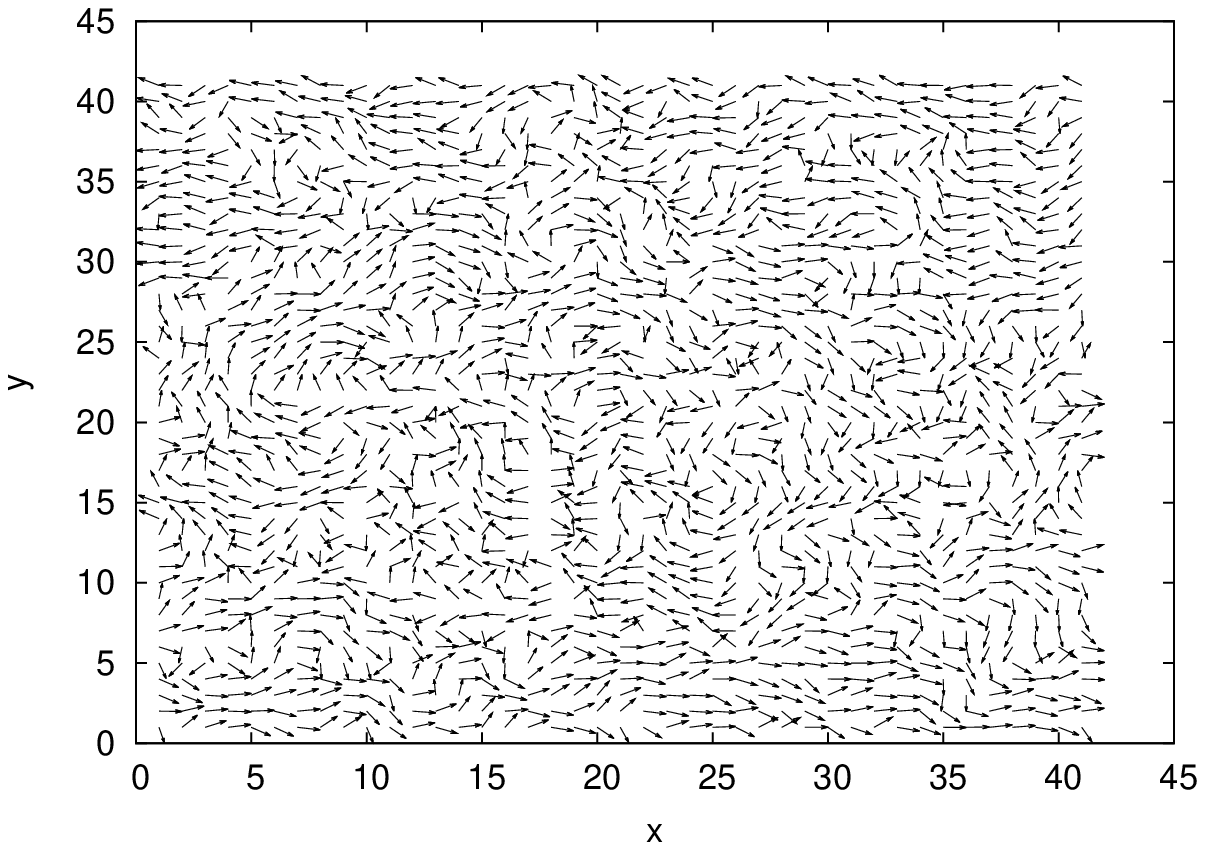}
\caption{
The normalized electromagnetic field ${\bf E}$ (left) and the polarization (right) for 
the Dirac model on a lattice with $41\times41$ sites for a single realization
of an Ising-like random mass with $|m|=4$.
}
\label{fig:emfi}
\end{figure*}

\begin{figure*}[t]
\includegraphics[width=7cm,height=7cm]{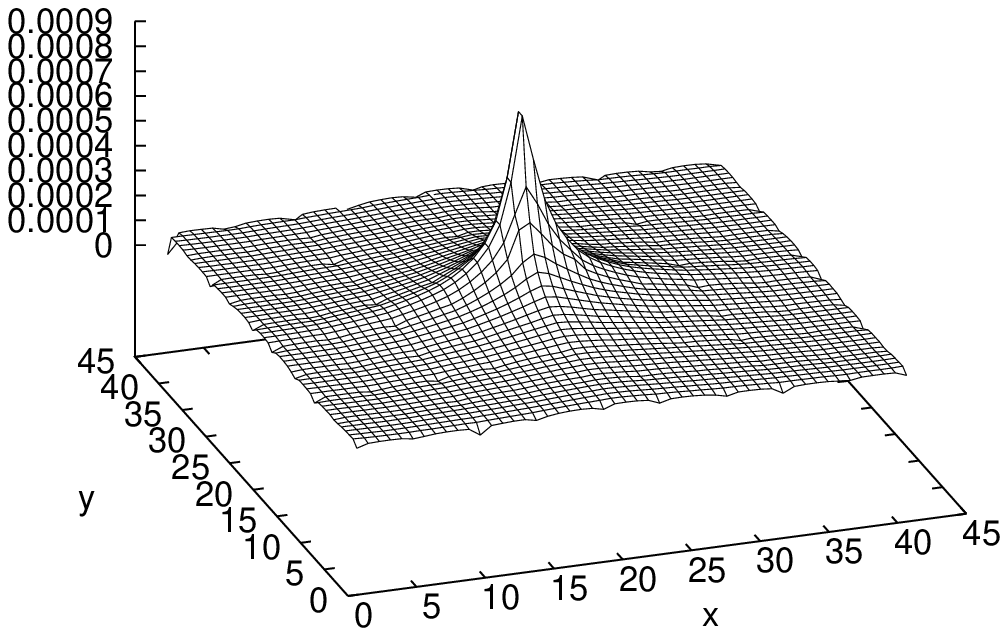}
\includegraphics[width=7cm,height=7cm]{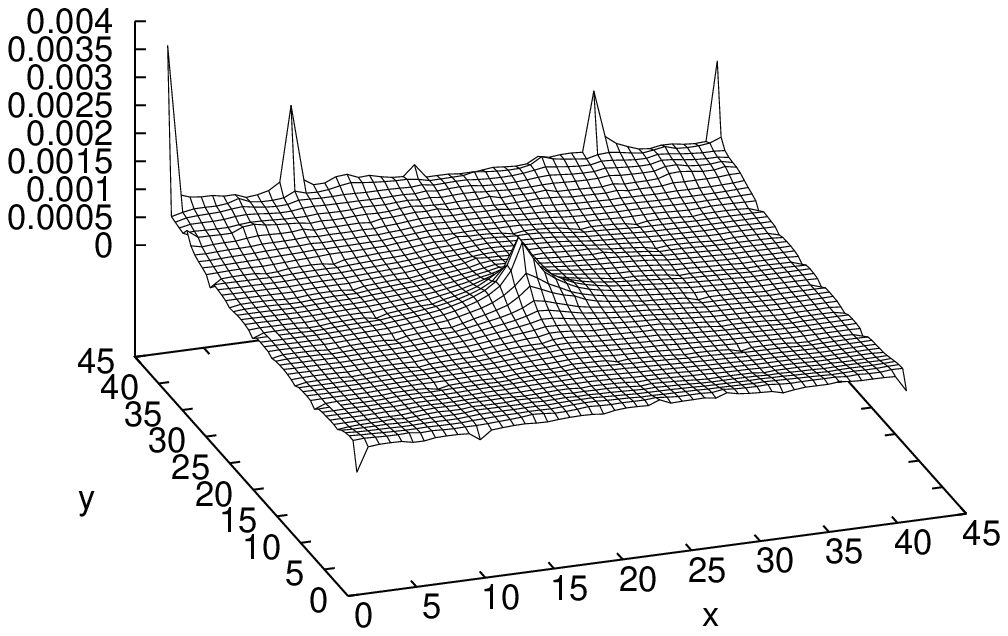}
\includegraphics[width=7cm,height=7cm]{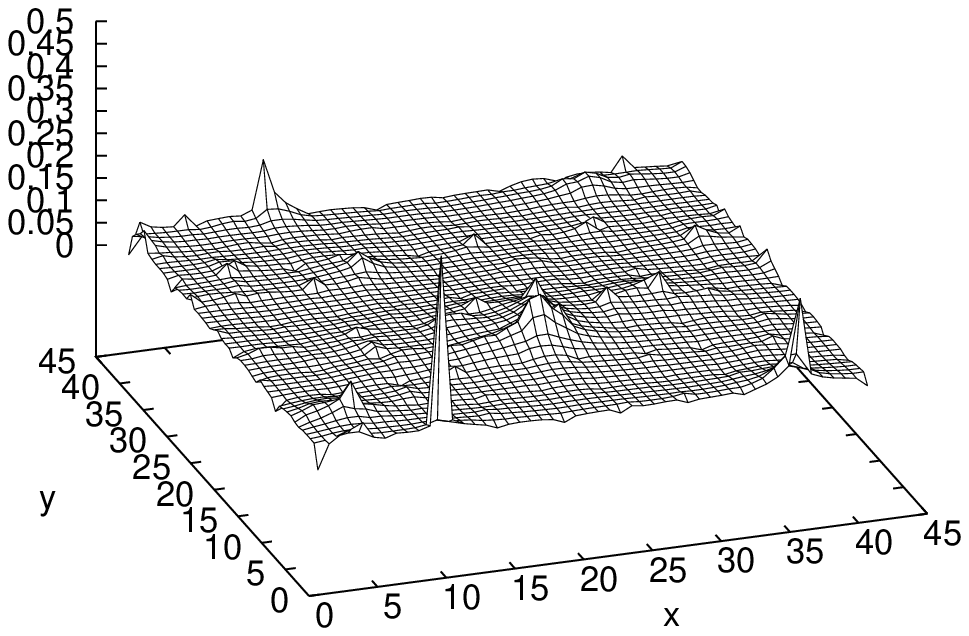}
\includegraphics[width=7cm,height=7cm]{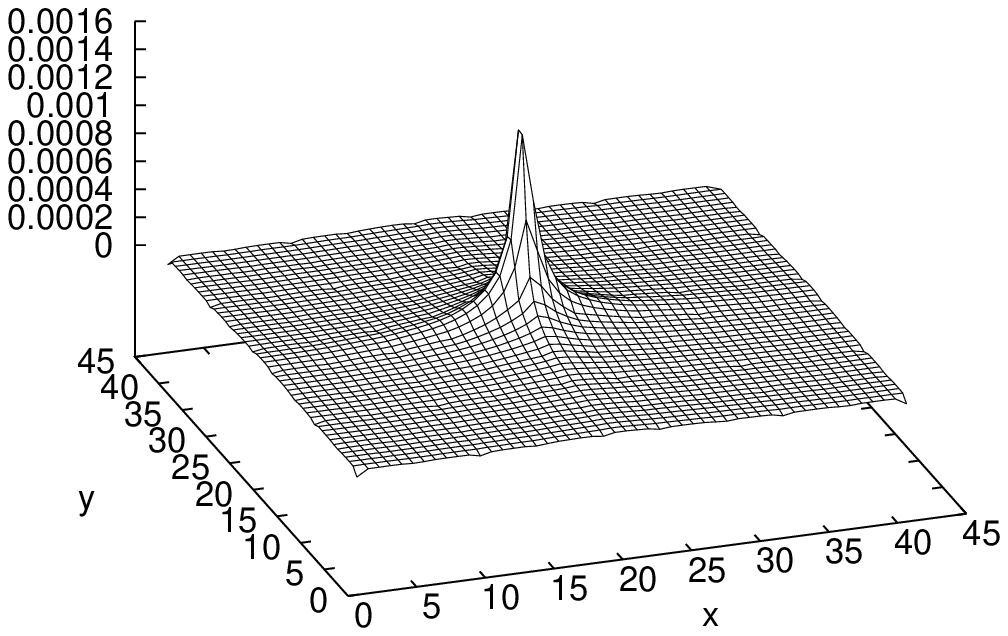}
\caption{
Intensity distribution from the central source for a single realization
of an Ising-like random mass with different values $|m|=0.5, 2, 10, 15$.
}
\label{fig:transfer1}
\end{figure*}

\subsection{Polarization of the electromagnetic field}

An electromagnetic field $\bE$ is characterized by four Stokes parameters \cite{hulst81,mishchenko06}.
They can be expressed as quadratic forms of the electric field $\bE$ with Pauli
matrices $\sigma_j$ ($j=0,x,y,z$; $\sigma_0$ is the $2\times 2$ unit matrix), where
$I=(\bE\cdot\sigma_0\bE)$ is the intensity with the scalar product $(.\cdot .)$ of
two-component vector field $\bE=(E_1,E_2)$ and 
\beq
Q=(\bE\cdot\sigma_z\bE)
\ , \ \ \ 
U=(\bE\cdot\sigma_x\bE)
\ , \ \ \ 
V=(\bE\cdot\sigma_y\bE)
\label{stokes00}
\eeq
are the other Stokes parameters which provide the polarization. 
It should be noticed that the relation $Q^2+U^2+V^2=I^2$ implies that the tip of
the vector $(Q,U,V)$ describes a sphere of radius $I$ (Poincar\'e sphere).
After normalizing the radius to 1, the vector field $(U,V,Q)/I$
can be used to characterize the polarization by a Berry curvature.

\section{Uniform gap}

First we consider the case of a spatially uniform Dirac mass. The resulting uniform gap in the spectrum
of $H_{DD}$ restricts the solution of Eq. (\ref{dirac01}) to an exponentially decaying field away
from the source on the length scale $l_m=1/\Delta_m$, where $\Delta_m$ is the effective gap created by
the uniform Dirac mass $m$.. If $l_m$ is larger (shorter) than the distance from
the source to the sample edge, an edge state is (not) excited. This is demonstrated in Fig. \ref{fig:uniform1},
where the source is either on the sample edge (left plot) or at the center (right plot), assuming that the
distance from the center to the edge is larger than $l_m$. Now we can reduce the gap to increase $l_m$
beyond the distance from the central source to the boundary. Then the field can reach the edge and an
edge state is created (cf. Fig. \ref{fig:uniform2}). Here we have used the small gap regimes of $H_{DD}$
for $m\approx 0$ and $m\approx 4$.

In Figs. \ref{fig:uniform1}, \ref{fig:uniform2} the intensity is plotted. Now we consider the electromagnetic
field and the polarization. Here and in the subsequent analysis, we assume that the source is always
at the center. The upper two plots in Fig. \ref{fig:uniform3} display the normalized field and the corresponding
polarization field, respectively, for $m=1$. The field at the source is $E_y=0$; i.e., the field
is oriented along the $x$ direction. Moreover, the polarization field is a vortex centered at the source
with a counterclockwise vorticity. Next we change the sign of $m$ globally to get $m=-1$ and obtain the two
plots at the bottom of Fig. \ref{fig:uniform3}: The sign change reverses the orientation of the electromagnetic
field and the vorticity of the polarization field, at least in the inner part of the sample. This effect is 
suppressed at boundaries in $y$-direction. Obviously, the edge states are much less affected by the sign change 
of $m$ than the field inside the sample.

\subsection{Straight internal edge}

A straight line along the $x$-axis, created by a sign flip in one half of the sample of an otherwise uniform
Dirac mass, leads to additional edge states~[\onlinecite{haldane08}]. If the source is located on this edge
there is a direct connection between the edge caused by the sign flip and the sample boundary. 
In this situation the field propagates along this line to the sample edge. The resulting intensity is 
visualized in Fig. \ref{fig:edge1}. The higher edge intensity at the boundary of the positive Dirac mass 
region is indicative of symmetry breaking due to a competition between different edge modes.

\begin{figure*}[t]
\includegraphics[width=7cm,height=7cm]{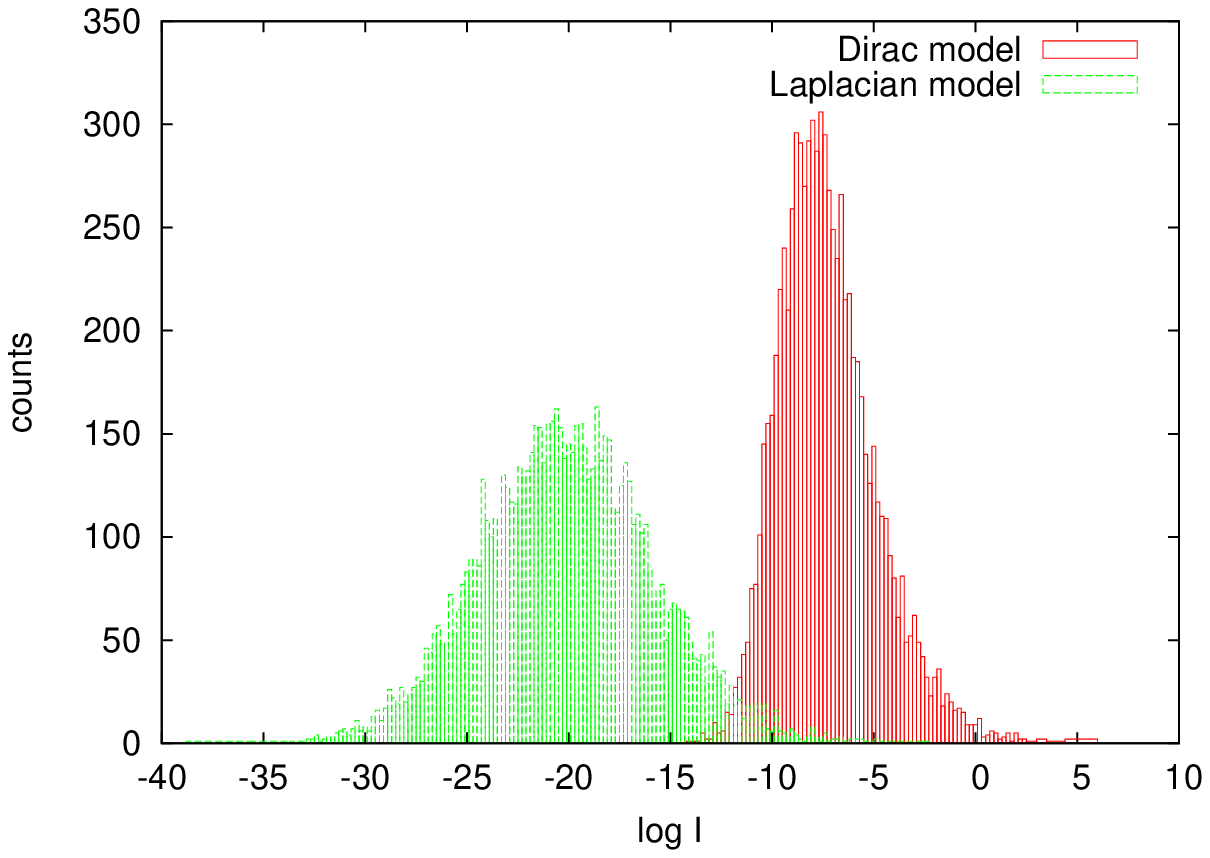}
\caption{
Lattice Dirac vs. Laplacian model: The intensity distribution is log-normal for the Laplacian model
and it is asymmetric for the Dirac model.
}
\label{fig:distr_hill1}
\end{figure*}

\begin{figure*}[t]
\includegraphics[width=7cm,height=7cm]{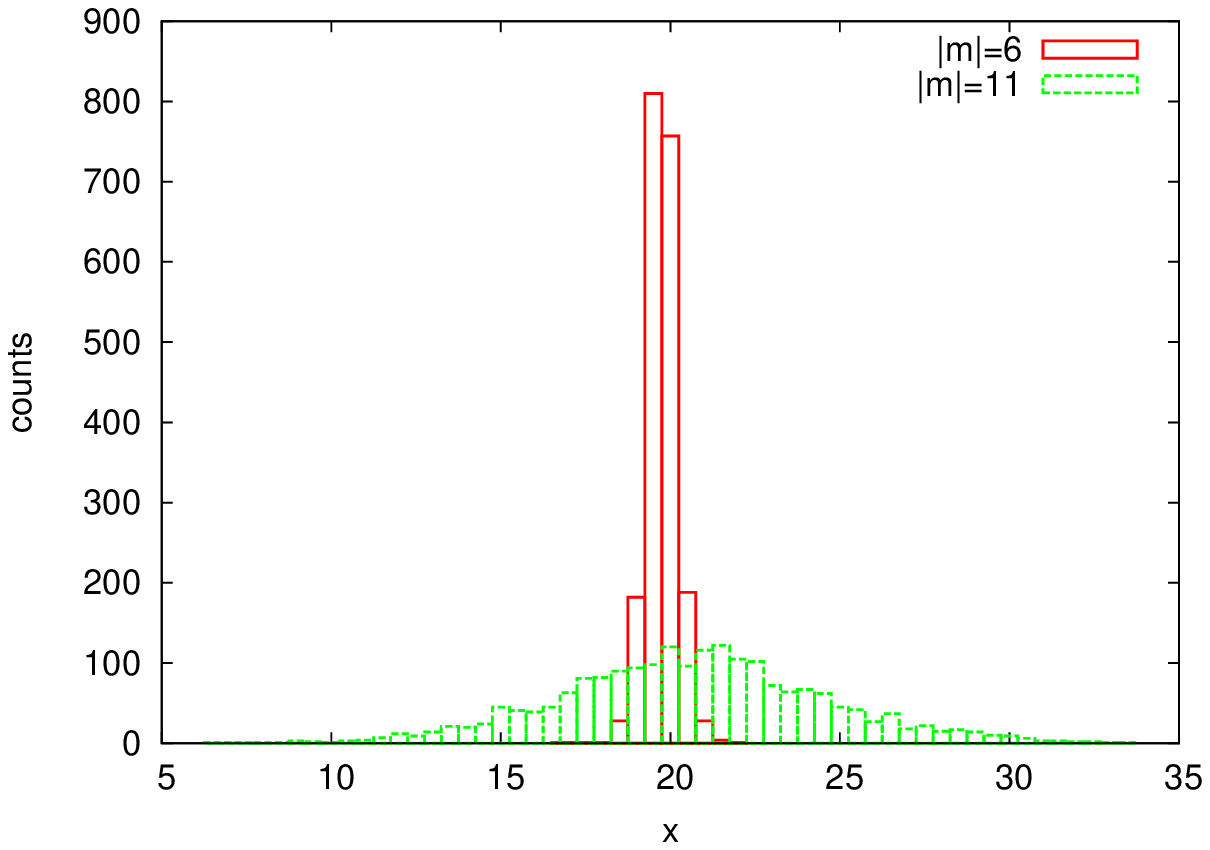}
\includegraphics[width=7cm,height=7cm]{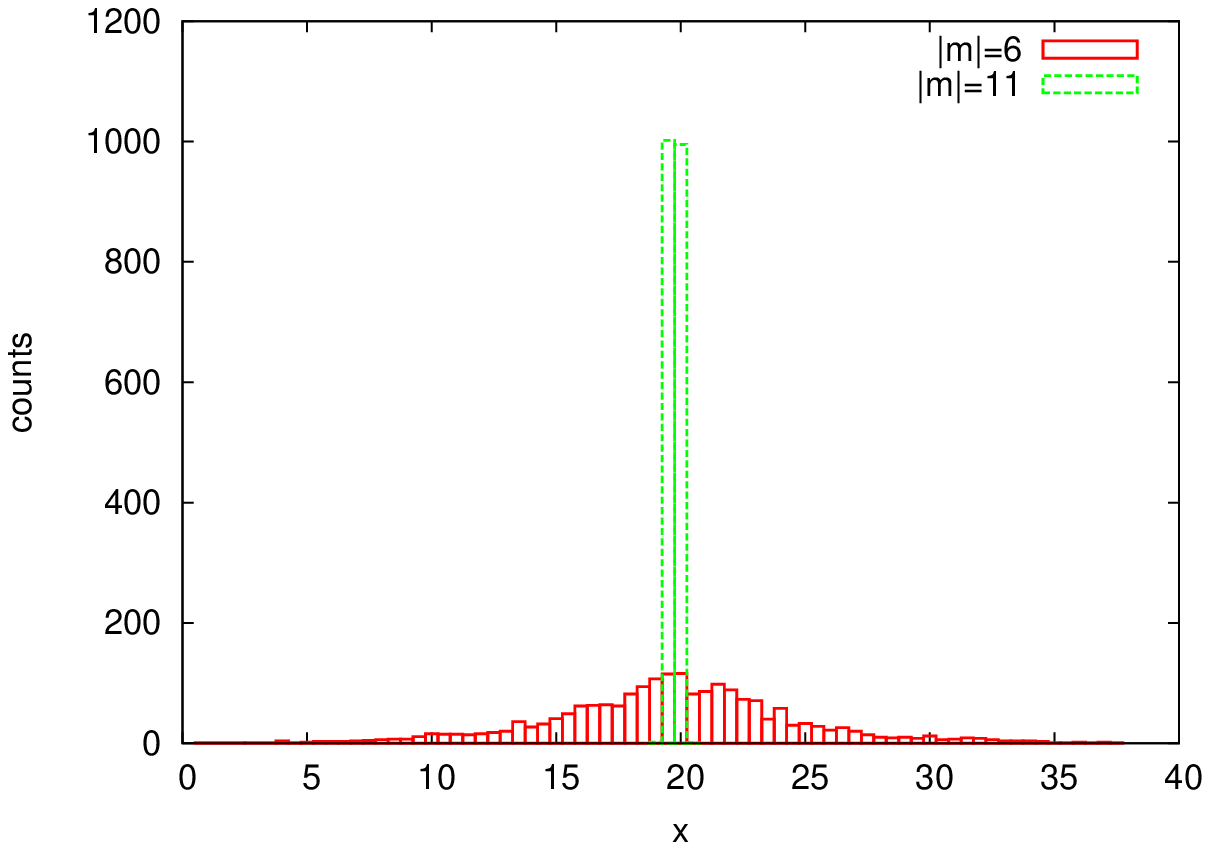}
\caption{
The distribution of $\langle x\rangle$ for the Dirac (left) and the Laplacian (right) model
on a $41\times 41$ lattice.
}
\label{fig:distr_hill2}
\end{figure*}

\begin{figure*}[t]
\includegraphics[width=7cm,height=7cm]{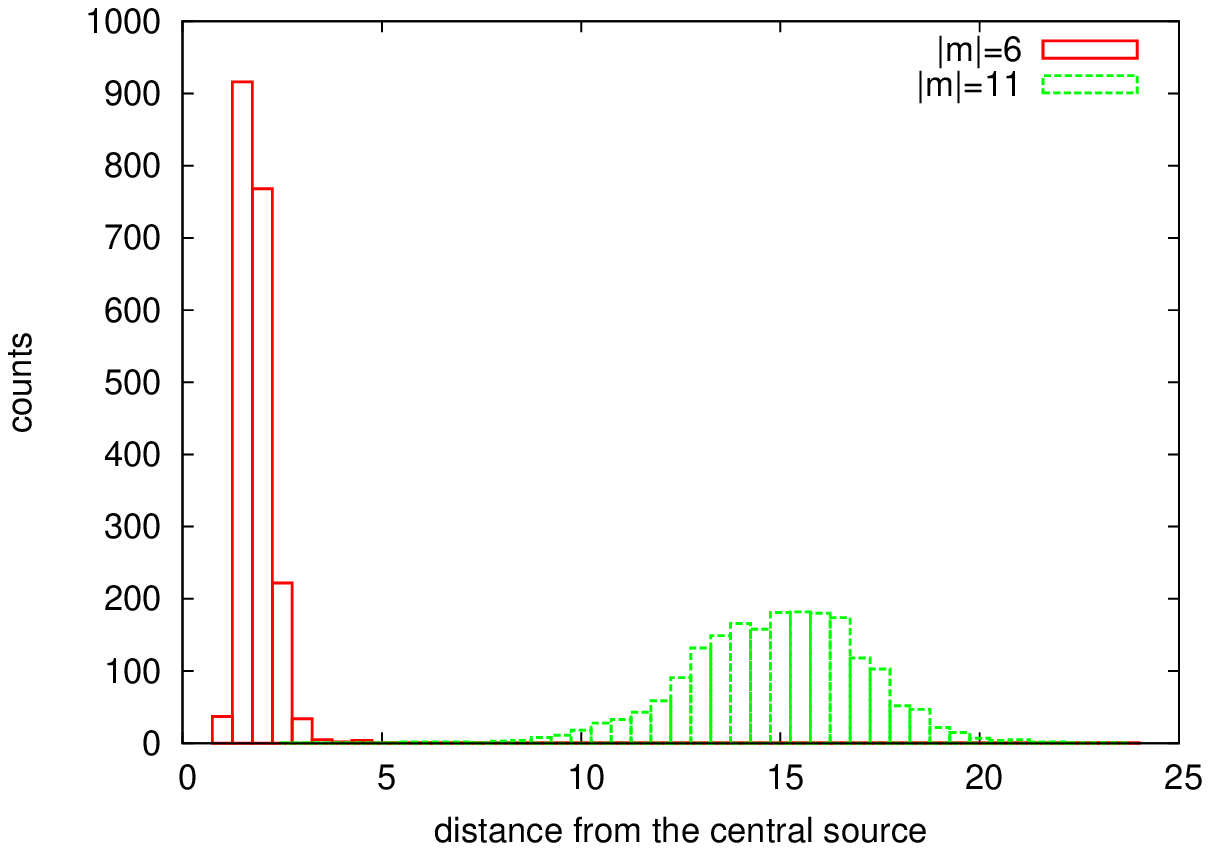}
\includegraphics[width=7cm,height=7cm]{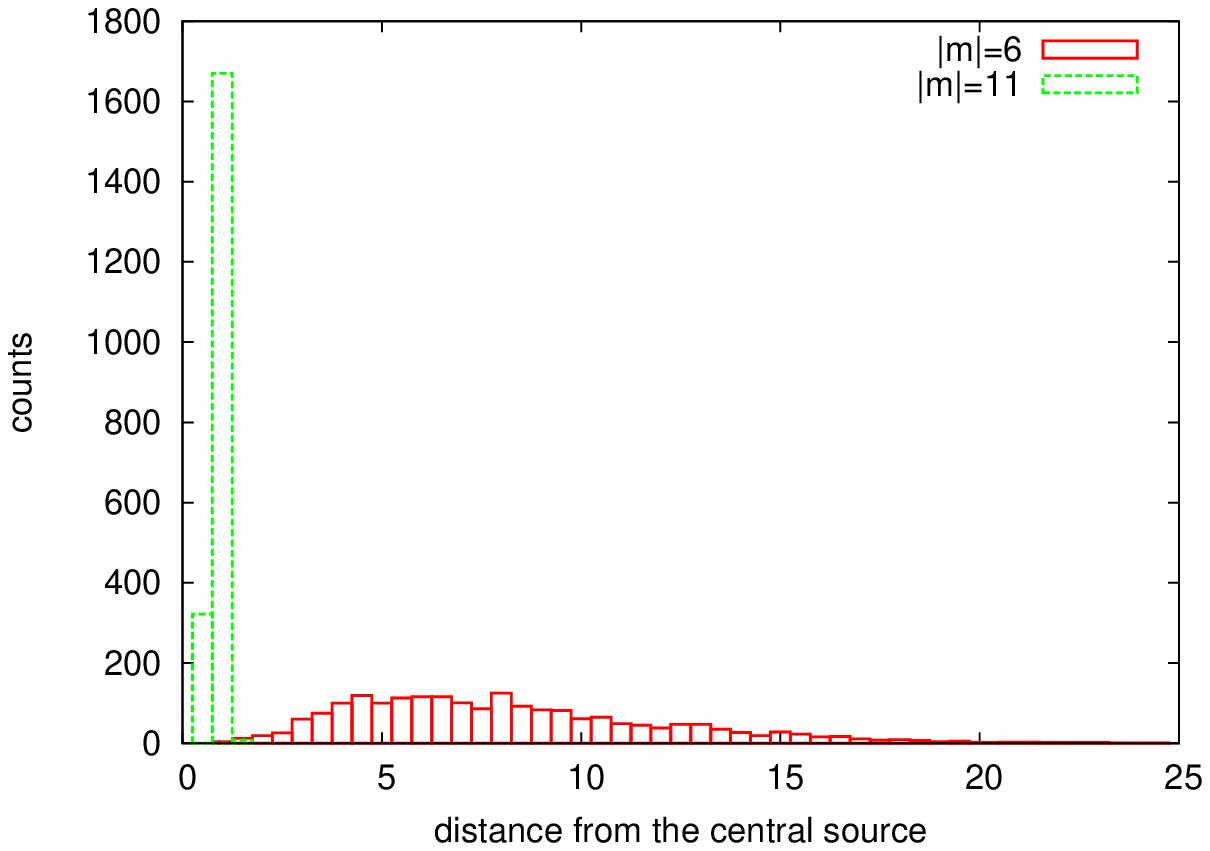}
\caption{
The distribution of the average distance from the central 
source for the Dirac model (left) and for the Laplacian model (right)
on a $41\times 41$ lattice.
}
\label{fig:distr_hill3}
\end{figure*}

\begin{figure*}[t]
\includegraphics[width=7cm,height=7cm]{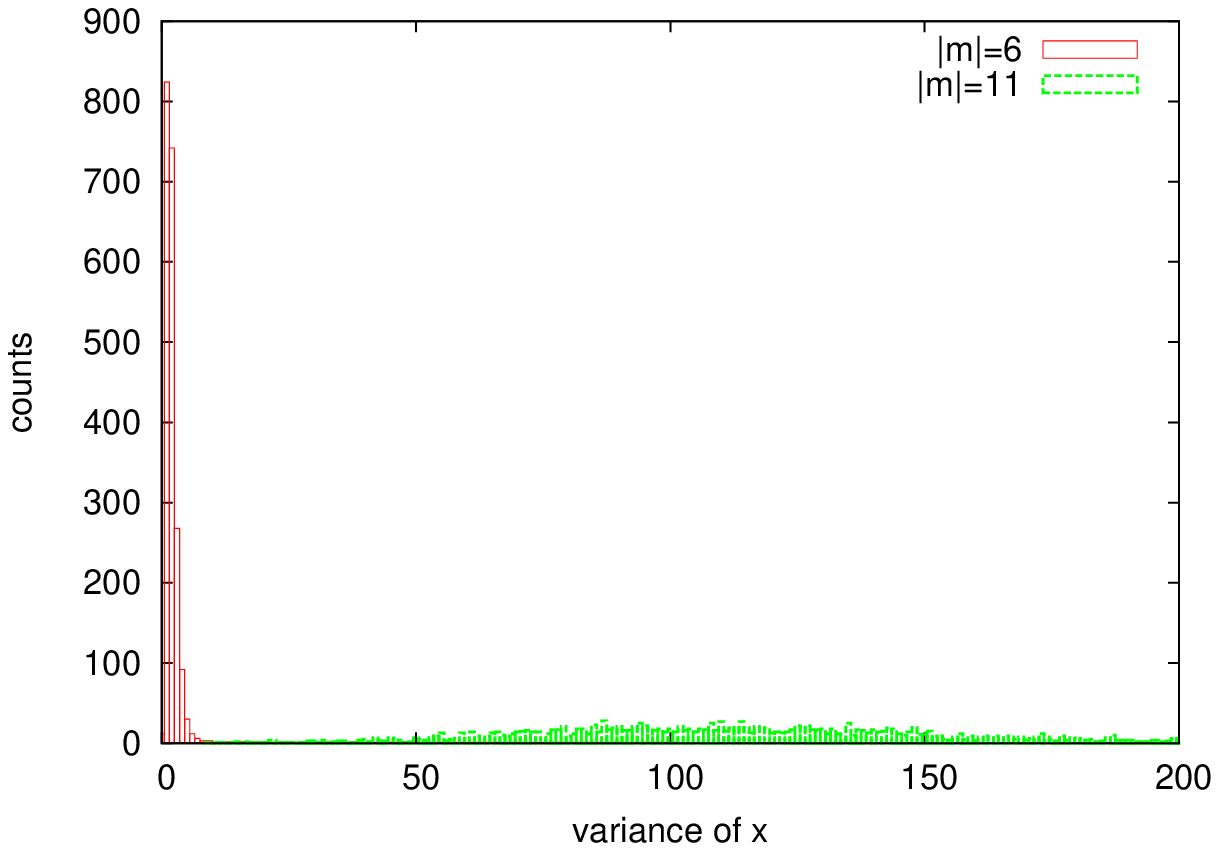}
\includegraphics[width=7cm,height=7cm]{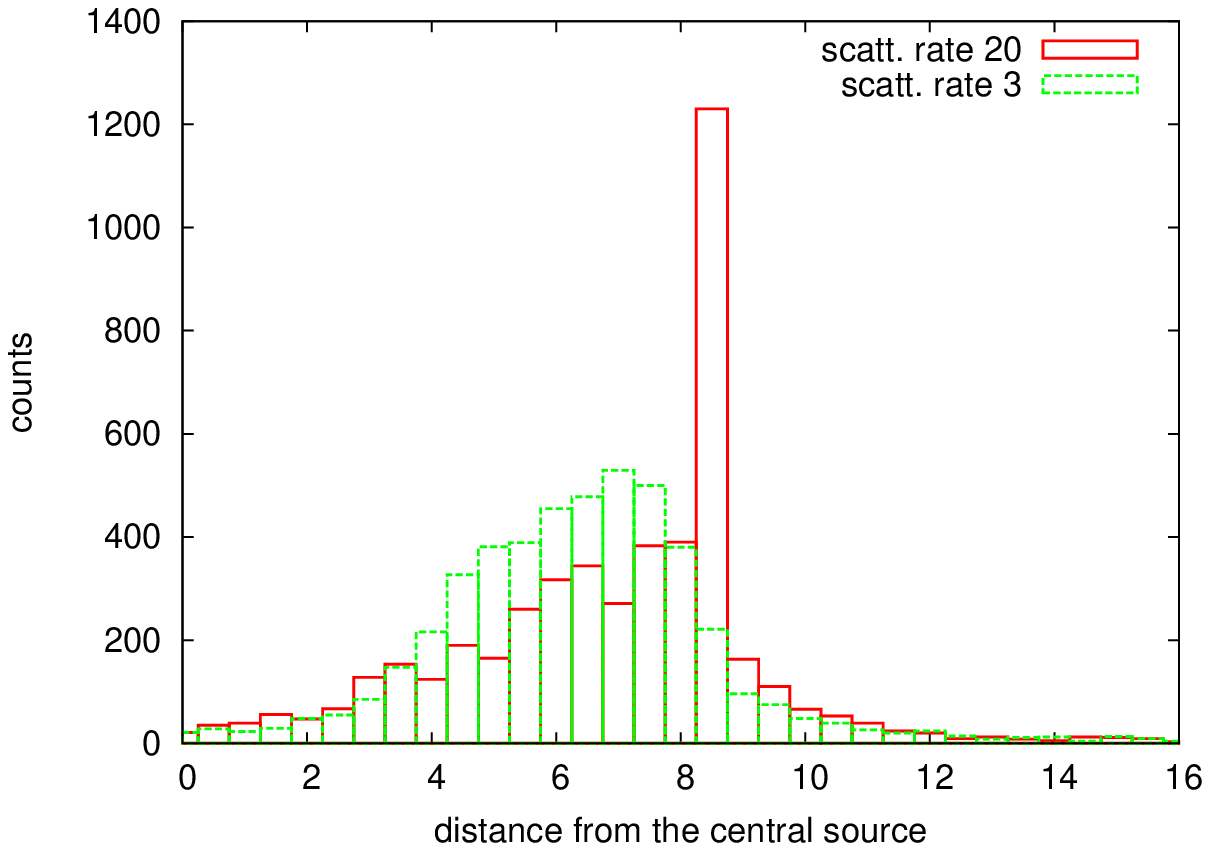}
\caption{
Dirac model: Distributions of the variance $C_{xx}$ for an 
Ising-like random Dirac mass on a $41\times 41$ lattice with $|m|=6$ and $|m|=11$ (left).\\
Invariant measure: The distribution of the average distance from the central source
on a $11\times 11$ lattice for two different scattering rates. 
The broadening of the distribution with increasing scattering rate is visible (right).
}
\label{fig:distr_hill4}
\end{figure*}

\section{Random Dirac mass}


When we assume that the Dirac mass $m$ is an uncorrelated spatial random number $m(\br)$, 
the resulting electromagnetic field as a solution of Eq. (\ref{dirac01}) and its polarization 
are also random. A single realization for a randomly fluctuating sign of the Dirac mass
is depicted in Fig. \ref{fig:emfi}. Here again we consider only
a central source in order to avoid a direct effect by the edge mode of sample boundary.
The random Dirac mass creates an ensemble of random edges between regions with
positive and negative
Dirac masses, respectively, which implies that the electromagnetic field can leave
the central source and reach the boundary by following these edges. However,
there is interference between the different edge states.
This could cause Anderson localization, which would
lead to an exponential decay of the intensity away from the central source.
But since large gap values lead to narrow edge states, the interference between 
different edges is also exponentially suppressed. Therefore, strong Dirac mass fluctuations 
should avoid conventional Anderson localization. Indeed,
previous studies for an infinite systems of random edges have revealed that
the intensity of an electromagnetic field can leave the source in the form of ray modes rather 
than being exponentially trapped at the source due to Anderson 
localization~[\onlinecite{1404,ziegler17a,ziegler17b}]. 
These ray modes obey a Fokker-Planck equation. Such a description cannot be directly
employed for finite systems because the reflexion by the boundary and the interaction
with the boundary edge state plays a significant role.

A more direct way to study the effect of edges is based on a numerical analysis of the electromagnetic
field in such a random environment. Some examples are given in Fig. \ref{fig:transfer1},
which confirm the behavior for strong randomness: While for weak randomness ($|m|=0.5$)
the field is localized around the source, for stronger randomness (e.g., for $|m|=10$)
the field is distributed over the entire lattice. Only for $|m|=15$ the field is localized
again around the source, since there are no edge states anymore due to the finite band width. 

A single realization of the random Dirac mass is too specific, though, to provide a predictive theoretical 
description and to compare with a realistic measurement. Therefore, it is important
to describe the random system by the distribution of physical quantities, such as the intensity. 
First, we consider the distribution of the intensity $I(\br)$ at site $\br$ in
Fig. \ref{fig:distr_hill1}. 
In this context it is also interesting to compare the electromagnetic field created with the Dirac equation
with that of the more conventional Helmholtz equation, where $H_{DD}$ is replaced by the 
lattice Laplacian $\Delta_1+\Delta_2-4+m$:
While for the gapless random Laplacian the intensity has a log-normal distribution, the 
logarithm of the intensity for $H_{DD}$ has an asymmetric distribution. 
This indicates that the intensity consists approximately of a product of independent random
variables only for the Laplacian model. 

Since the intensity is a function of the coordinates $\br=(x,y)$, it can be used to define
\beq
P_I(\br)=\frac{I(\br)}{\sum_{\br}I(\br)}
\ ,
\label{int_distr}
\eeq
which is a probability measure for the normalized intensity at the gives position $(x,y)$.
With $P_I(\br)$ we can evaluate expectation values of the coordinates such as
\beq
\langle f(\br)\rangle
=\sum_{\br}f(\br)P_I(\br)
\eeq
for {\it each} realization of the random Dirac mass. For instance, $\langle \br\rangle$ reveals 
approximately the position of the highest intensity in space (i.e., the center of intensity). 
A related quantity is the mean square displacement (or variance) $C_{xx} 
=\langle x^2\rangle-\langle x\rangle^2$,
which measures the strength of the fluctuations of the intensity with respect to the center of intensity. 
All these quantities characterize the intensity for a single realization of the random Dirac mass.
This is relevant for a single experiment, in which we choose a certain realization of
the photonic metamaterial. However, it is also of interest how a certain distribution of the Dirac mass
(for instance, an Ising-like $\pm m$ or a Gaussian distribution) would create a distribution
of these quantities. For this purpose we consider many realizations
of the Dirac mass to obtain the (un-normalized) distribution $\Pi(\langle \br\rangle)$ and 
related distributions of $C_{xx}$.
The latter allows us to determine the presence of Anderson localization, for which
the distribution would be strongly peaked at the position of the source and decay
exponentially away from it. A broadening of the distribution for increasing randomness, on the other hand,
indicates anti-localization. In such a case the  random edges support the propagation of 
the electromagnetic field away from the source.

Both effects can be seen in Figs. \ref{fig:distr_hill2} -- \ref{fig:distr_hill4}, where we compare
different distributions generated from the lattice Dirac model and from the random Laplacian model.
$m$ has a fluctuating sign and the values $|m|=6$ and $|m|=11$ are compared.
The distribution of $\langle x\rangle$ is plotted in Fig. \ref{fig:distr_hill2} and 
the distribution for the average distance from the central source is depicted in Fig. \ref{fig:distr_hill3}.
Finally, the distribution of $C_{xx}$ (left plot of Fig. \ref{fig:distr_hill4}) 
is analyzed.
Obviously, increasing randomness caused by an increasing $|m|$ leads to a broadening (narrowing)
of the considered distributions in the case of the lattice Dirac (Laplacian) model, respectively.
It is remarkable that the two values of $|m|$ lead to very distinct distributions.
They are very narrow around the source for $|m|=6$ but very broad for $|m|=11$ for the lattice Dirac model
and vice versa for the Laplacian model. Moreover, the variance $C_{xx}$
vanishes quickly away from zero for $|m|=6$ but is very broad for $|m|=11$ in the case of the lattice 
Dirac model. On the other hand, the distribution of $\langle x\rangle$ in the case of the Laplacian model is 
broad for $|m|=6$ but very narrow for $|m|=11$ (right plot of Fig. \ref{fig:distr_hill2}). This is 
a clear signature of Anderson localization.

\section{Discussion}

The examples in Fig. \ref{fig:uniform3} demonstrate the sensitivity of the
electromagnetic field and its polarization under a sign change of a uniform 
Dirac mass in a finite system. A positive Dirac mass together with the boundary
enforces a vortex for the polarization field with a counter-clockwise vorticity. 
On the other hand, a negative uniform Dirac mass creates a vortex with clockwise vorticity, 
which is suppressed by the edge state at the boundary.
A similar competing effect between a negative Dirac mass and the edge state exists for 
the two-component electromagnetic field. This indicates that the boundary has a strong
influence on the properties of the field.

In general, the electromagnetic field can consist of edge states
that are created by the interfaces between regions of positive and negative masses 
(cf. Fig. \ref{fig:edge1}). 
This is also the case for a randomly fluctuating mass sign in the Dirac model, as demonstrated
for the normalized electromagnetic field and its polarization field in Fig. \ref{fig:emfi}. 
These edge states defeat Anderson localization and enable the field to spread in space with
the consequence that the field intensity is distributed over the entire finite
system. This effect even increases with increasing randomness, which reflects the fact that
a larger $|m|$ creates sharper edge states to avoid a decay of the intensity due to destructive 
interference.

In contrast to the absence of Anderson localization in the lattice Dirac model
there is a tendency in the Laplacian model to Anderson localization with increasing randomness
(cf. Figs. \ref{fig:distr_hill2}, \ref{fig:distr_hill3}). The latter model has 
only a single band and no edge states, which implies a conventional localization 
behavior~[\onlinecite{anderson58,abrahams79,wegner79}]. In this context it could be interesting
to consider the cross-over operator
\[
H_{DD}'=\frac{v_D}{a}\pmatrix{
m +\Delta_1+\Delta_2-4 & \beta(id_1+d_2) \cr
\beta(id_1-d_2) & -m-\Delta_1-\Delta_2+4 \cr
}
\ ,
\]
where the parameter $\beta$ can be tuned between 0 and 1. This offers an opportunity to study
the transition to Anderson localization by approaching $\beta=0$.

In this work we have only presented results for a randomly fluctuating sign of the
Dirac mass (i.e., Ising-like randomness). At least qualitatively the results are also
valid for a continuous distribution of the Dirac mass, such as a Gaussian distribution.
From this point of view the absence of Anderson localization is generic for the random mass
Dirac model.

\subsection{Note on the random phase model}

The distribution of the local intensity $I(\br)$ was studied in Refs.~[\onlinecite{1404,ziegler17a,ziegler17b}],
using its invariant measure. The latter is associated with an internal global symmetry of the
distribution, which creates long-range correlations. The invariant measure can be understood as a projection
which eliminates all fluctuations not invariant under the symmetry transformation. Then the strength of the 
random fluctuations is expressed by an effective scattering rate that is a monotonic function of the variance
of the random fluctuations. This concept is fundamental
to the Random Matrix Theory~[\onlinecite{mehta91}] and leads to a random phase model in the case
of the random mass Dirac model~[\onlinecite{1404}].
The invariant measure corresponds to the Jacobian determinant related to the symmetry degrees of freedom.
A graphical representation of a determinant on the lattice consists of non-intersecting loops. Then the random 
phase factors create vertices with four legs along these loops~[\onlinecite{1404}]. Thus, at most two 
(complex conjugate) phase factor appear at a given lattice site. This enables us to replace the continuous random phase 
by a discrete random phase with the two values $\{0,\pi\}$, which is equivalent to replacing the
phase factors by Ising spins. This step simplifies the numerical simulation substantially. 
Although the invariant measure covers only the long-range asymptotics of the random intensity distribution, there 
is also an increasing transfer of intensity from the central source to the boundary with increasing
randomness for the finite system (right plot in Fig. \ref{fig:distr_hill4}), a behavior reminiscent of the 
Dirac model with random mass.

\section{Conclusion}

Both, the electromagnetic field and its polarization are strongly affected by the sign of the
Dirac mass and by the edge state at the sample boundary. 
The numerical simulation of a finite two-dimensional lattice Dirac model with random mass
demonstrates that the ensemble of edge states leads to a broad spatial intensity distribution. 
Using an Ising-like fluctuating sign of the Dirac mass, the width of the intensity distribution 
broadens with increasing $|m|$. In contrast, randomness in a one-band model has the opposite effect,
where the intensity distribution narrows down around the source with increasing $|m|$.
This indicates Anderson localization. 

\vskip0.3cm

\no
Acknowledgment: This work was supported by a grant of the Julian Schwinger Foundation.


\begin{thebibliography}{99}


\bibitem{maradudin91}
M. Plihal and A.A. Maradudin, Phys. Rev. B {\bf 44}, 8565 (1991).

\bibitem{peleg07}
O. Peleg, et al.,
Phys. Rev. Lett. {\bf 98}, 103901 (2007). 

\bibitem{haldane08}
F.D.M. Haldane and S. Raghu, Phys. Rev. Lett. {\bf 100}, 013904 (2008).

\bibitem{raghu08}
S. Raghu and F.D.M. Haldane, Phys. Rev. A {\bf 78}, 033834 (2008).

\bibitem{zhang08}
X. Zhang, Phys. Rev. Lett. {\bf 100}, 113903 (2008).

\bibitem{ochiai09}
T. Ochiai and M. Onoda, Phys. Rev. B {\bf 80}, 155103 (2009)

\bibitem{ablowitz09}
M.J. Ablowitz, S.D. Nixon, and Y. Zhu, 
Phys. Rev. A {\bf 79}, 053830 (2009). 

\bibitem{wang09}
L.G. Wang, Z.G. Wang, J.X. Zhang and S.Y. Zhu, Opt. Lett. {\bf 34}, 1510-1512 (2009).

\bibitem{zandbergen10}
S.R. Zandbergen and M.J.A. de Dood, Phys. Rev. Lett. {\bf 104}, 043903 (2010).

\bibitem{huang11}
X. Huang, Y. Lai, Z. H. Hang, H. Zheng and C.T. Chan, Nature Materials {\bf 10}, 582-586 (2011).

\bibitem{bravo12}
J. Bravo-Abad, J.D. Joannopoulos and M. Solja\v{c}i\'c, PNAS {\bf 109}, 9761 (2012).

\bibitem{fefferman12}
C.L. Fefferman and M.I. Weinstein, J. Amer. Math. Soc. {\bf 25}, 1169-1220 (2012).

\bibitem{rechtsman13}
M.C. Rechtsman et al.,
Nature {\bf 496}, 196 (2013).

\bibitem{keil13}
R. Keil et al. 
Nat. Commun. 4:1368 (2013). 

\bibitem{ma15}
T. Ma, A.B. Khanikaev, S. Hossein Mousavi, and G. Shvets, Phys. Rev. Lett. {\bf 114}, 127401 (2015).

\bibitem{cheng16}
X. Cheng et al., 
Nature Mater. {\bf 15}, 542-548 (2016).

\bibitem{klitzing80}
K. v. Klitzing, G. Dorda, and M. Pepper, Phys. Rev. Lett. {\bf 45}, 494 (1980).

\bibitem{1404}
K. Ziegler,  J. Phys. A: Math. Theor. {\bf 48}, 055102 (2015).

\bibitem{ziegler17a}
K. Ziegler, Journ. Phys. A: Math. and Theor. Phys. {\bf 50}, 125002 (2017).
%
\bibitem{ziegler17b}
K. Ziegler, Ann. Phys. {\bf 529} (8), 1600345 (2017).

\bibitem{john87}
S. John, Phys. Rev. Lett. {\bf 87}, 2486 (1987).

\bibitem{bykov72}
V.P. Bykov, 
Soviet Journal of Experimental and Theoretical Physics {\bf 35}, 269 (1972).

\bibitem{ohtaka87}
K. Ohtaka, 
Phys. Rev. B {\bf 19}, 5057 (1979).

\bibitem{yablonovitch87}
E. Yablonovitch, Phys. Rev. Lett. {\bf 58}, 2043 (1987).

\bibitem{Susskind1977}
L. Susskind, Phys. Rev. D {\bf 16}, 3031 (1977).

\bibitem{beenakker08}
J. Tworzyd\l o, C.W. Groth and C.W.J. Beenakker, Phys. Rev. B {\bf 78}, 235438 (2008).

\bibitem{Stacey1982}
R. Stacey, Phys. Rev. D {\bf 26}, 468 (1982).

\bibitem{ziegler96}
K. Ziegler, Phys. Rev. B {\bf 53}, 9653 (1996).

\bibitem{beenakker10}
M. V. Medvedyeva, J. Tworzyd\l o, and C. W. J. Beenakker, Phys. Rev. B {\bf 81}, 214203 (2010).

\bibitem{hill14}
A. Hill and K. Ziegler, Eur. Phys. J. B {\bf 87}, 142 (2014). 

\bibitem{hulst81}
H.C. van de Hulst, {\it Light Scattering by Small Particles}, Dover Publications
(New York 1981).

\bibitem{mishchenko06}
M.I. Mishchenko, L.D. Travis and A.A. Lacis, {\it Multiple scattering of light by particles},
Cambridge University Press (Cambridge 2006).

\bibitem{anderson58}
P.W. Anderson, Phys. Rev. {\bf 109}, 1492 (1958).

\bibitem{abrahams79}
E. Abrahams, P.W. Anderson, D.C. Licciardello and T.V. Ramakrishnan,
Phys. Rev. Lett. {\bf 42}, 673 (1979).

\bibitem{wegner79}
F. Wegner, Z. Phys. B {\bf 35}, 207 (1979).

\bibitem{mehta91}
M.L. Mehta, {\it Random Matrices} (Academic Press 1991). 

\end{thebibliography}
\end{document}